\documentclass[11pt]{article} 

\usepackage[utf8]{inputenc} 
\usepackage{amsmath}

\usepackage{epstopdf}

\usepackage{geometry} 
\usepackage{graphicx} 
\usepackage[affil-it]{authblk}

\usepackage{booktabs} 
\usepackage{array} 
\usepackage{paralist} 
\usepackage{verbatim} 
\usepackage{subfig} 

\usepackage{fancyhdr} 
\usepackage{sectsty}
\allsectionsfont{\sffamily\mdseries\upshape} 

\usepackage[nottoc,notlof,notlot]{tocbibind} 
\usepackage[titles,subfigure]{tocloft} 

\title{Diverse fluctuations and anisotropic Gr{\" u}neisen parameter behavior in iron-based superconductor Ba(Fe$_{1-x}$Co$_x$)$_2$As$_2$
and their correlation with superconductivity}

\author{Chiaki Fujii$^{1}$, Shalamujiang Simayi\thanks{Present address: Renewable Energy Research Center, AIST}$^{1}$, Kouhei Sakano$^{1}$, Chizuru Sasaki$^{1}$, Mitsuteru Nakamura$^{1}$, Yoshiki Nakanishi$^{1,4}$, Kunihiro Kihou$^{2,4}$, 
Masamichi Nakajima\thanks{Present address: Faculty of Science, Osaka University}$^{2,4}$,
Chul-Ho Lee$^{2,4}$,
Akira Iyo$^{2,4}$, Hiroshi Eisaki$^{2,4}$, 
Shin-ichi Uchida$^{2,3,4}$, and
Masahito Yoshizawa\thanks{E-mail address: yoshizawa@iwate-u.ac.jp}$^{1,4}$
}

\affil{$^{1}$ Graduate School of Engineering, Iwate University, Morioka 020-8551\\
$^{2}$ National Institute of Advanced Industrial Science and Technology (AIST), Tsukuba 305-8568 \\
$^{3}$ Graduate School of Science, The University of Tokyo, Tokyo 113-0033 \\
$^{4}$ Transformative Research-project on Iron Pnictides (TRIP), Japan Science and Technology Agency, Tokyo 102-0075}

\begin{document}
\maketitle
\abstract{
In this study, the temperature dependence of elastic constants $C_{11}$, $C_{33}$, $C_{\rm E} = (C_{11}-C_{12})/2$, $C_{66}$ and $C_{44}$ of the iron-based superconductor Ba(Fe$_{1-x}$Co$_{x}$)$_{2}$As$_{2}$ (0 to 0.245) have been measured. 
This system shows a large elastic softening in $C_{66}$ towards low temperatures.
In addition to $C_{66}$, which originates from orthorhombic structural fluctuation, the samples near the optimal concentration show remarkable structural fluctuation in $C_{11}$ and $C_{33}$ elastic modes, which correspond to $\Gamma_{1}$ (C4) symmetry. It suggests the existence of diverse fluctuations in this system. Gr{\" u}neisen parameters were analyzed under some assumptions for structural and magnetic transition temperatures. Results showed that the Gr{\" u}neisen parameters for the inter-plane strain are remarkably enhanced toward the QCP, while those for the in-plane stress tend to turn down near the QCP. Gr{\" u}neisen parameters for the superconducting transition are anisotropic and shows remarkable Co-concentration dependence, suggesting that the in-plane isotropic compression and inter-layer elongation enhance the superconductivity. The correlation of Gr{\" u}neisen parameters between $T_{\rm S}$, $T_{\rm N}$ and $T_{\rm sc}$ shows $c$-axis elongation and its relevant role in the emergence of superconductivity in this system.
}



\section{Introduction}
Today, superconductors are widely used in daily life. 
Magnetic resonance imaging (MRI) working with superconducting magnet provides with high resolution image of human body, and a powerful tool for medical diagnosis.
Superconducting quantum interference devices (SQUID) has been used for brain research and detection of diseases.
Novel superconductors (SCs) with higher critical temperature $T_{\rm sc}$ will be desired for wider applications of SCs. 
In 2008, iron-based superconductor (SC) was discovered by Hosono Group at Tokyo Institute of Technology.\cite{kamihara2008}
This new type of SC has been attracted much attention and promoted projects due to its relatively high superconducting transition temperature $T_{\rm sc}$ although it contains a magnetic ion Fe as a constitutional element.\cite{johnston2010}
Such high $T_{\rm sc}$ of iron-based SCs has not been brought about by phonon, but possibly by other mechanisms.\cite{boeri2008}
Therefore, it is expected that investigation of iron-based materials would lead to a discovery of novel SCs with higher $T_{\rm sc}$.  

\begin{figure}
\begin{center}
\includegraphics[width=8cm]{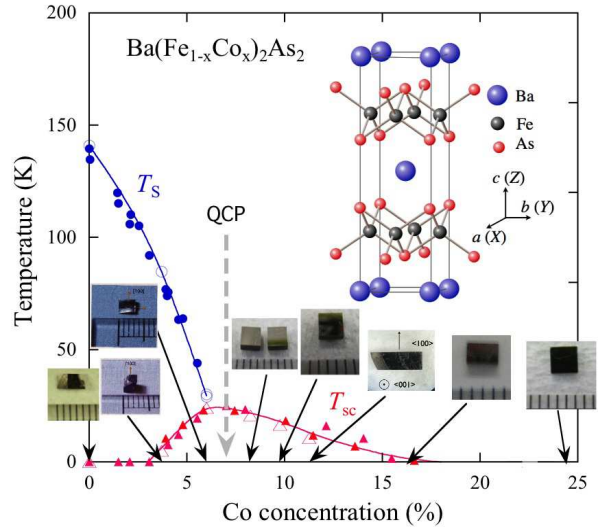}
\end{center}
\caption{(Color online) Phase diagram and crystal structure of Ba(Fe$_{1-x}$Co$_{x}$)$_{2}$As$_{2}$ . Crystal structure of BaFe$_2$As$_2$ belongs to the base-centered tetragonal crystal class $I4/mmm$. The adopted $XYZ$ coordinates are described in the figure. These pictures of eight single crystals used in this experiment with their locations in the phase diagram are also shown in the figure. $T_{\rm S}$ and $T_{\rm sc}$ are the structural and superconducting phase transition temperatures. Open symbols are obtained by the previous work.\protect\cite{yoshizawa2012-1} Closed symbols are reported in the previous studies.\protect\cite{canfield2009,laplace2009}   }
\label{sample}
\end{figure}

One of the key strategies to investigate the superconductivity is quantum criticality. 
According to recent studies for oxide SCs and strong correlated SCs, the superconducting phases of these systems are located near the neighboring phases like magnetic order.
These neighboring phases disappear towards quantum critical point (QCP) by tuning control parameters such as hydrostatic pressure, chemical doping and magnetic field, and superconductivity turns out.
Figure \ref{sample} shows the phase diagram of Ba(Fe$_{1-x}$Co$_{x}$)$_{2}$As$_{2}$ (Ba122), which will be reported in this article. As can be seen in the Fig. \ref{sample}, the structural transition temperature $T_{\rm S}$ of the parent compound BaFe$_{2}$As$_{2}$ is 134.4 K, and it gradually decreases by replacing Fe by Co. $T_{\rm S}$ gradually decreases to zero at the QCP (x=0.07), and where $T_{\rm sc}$ reached to the maximum value. 
These alternations of the phases such as structural, magnetic and superconducting one suggests the importance of the neighboring phase in the emergence of superconductivity in iron based superconducting system. 
Fluctuations of the neighboring phase possibly mediates superconductivity.
This consideration would lead us to find out a new type of superconductivity through the systematic investigations of the order and its fluctuation of the neighboring phase.
Iron-based SCs show magnetic and structural (orbital) phases near to the superconducting phase, therefore there are two proposals were objected for the superconducting mechanism of iron-based SCs, they are magnetic and orbital characters.
Namely, spin fluctuations and orbital fluctuations are new probable candidates for the superconductivity of iron-based SCs.

Investigation on the order of the neighboring phases and the order parameter fluctuations would be crucial for the superconductivity research.
Elastic constant measurements are very powerful tools to investigate structural fluctuations. Ultrasonic wave and elastic strain introduced into solid for the sound velocity measurements breaks the local symmetry of the crystal.
Because, as illustrated in Fig. 2, these elastic strains have the same symmetry as the elastic quadrupoles (rank-2 multipoles), therefore they can couple with the orbital degrees of freedom.
As a result, those elastic constants which are the strain susceptibilities provide the relevant informations on the orbitals.

\begin{figure}
\begin{center}
\includegraphics[width=8cm]{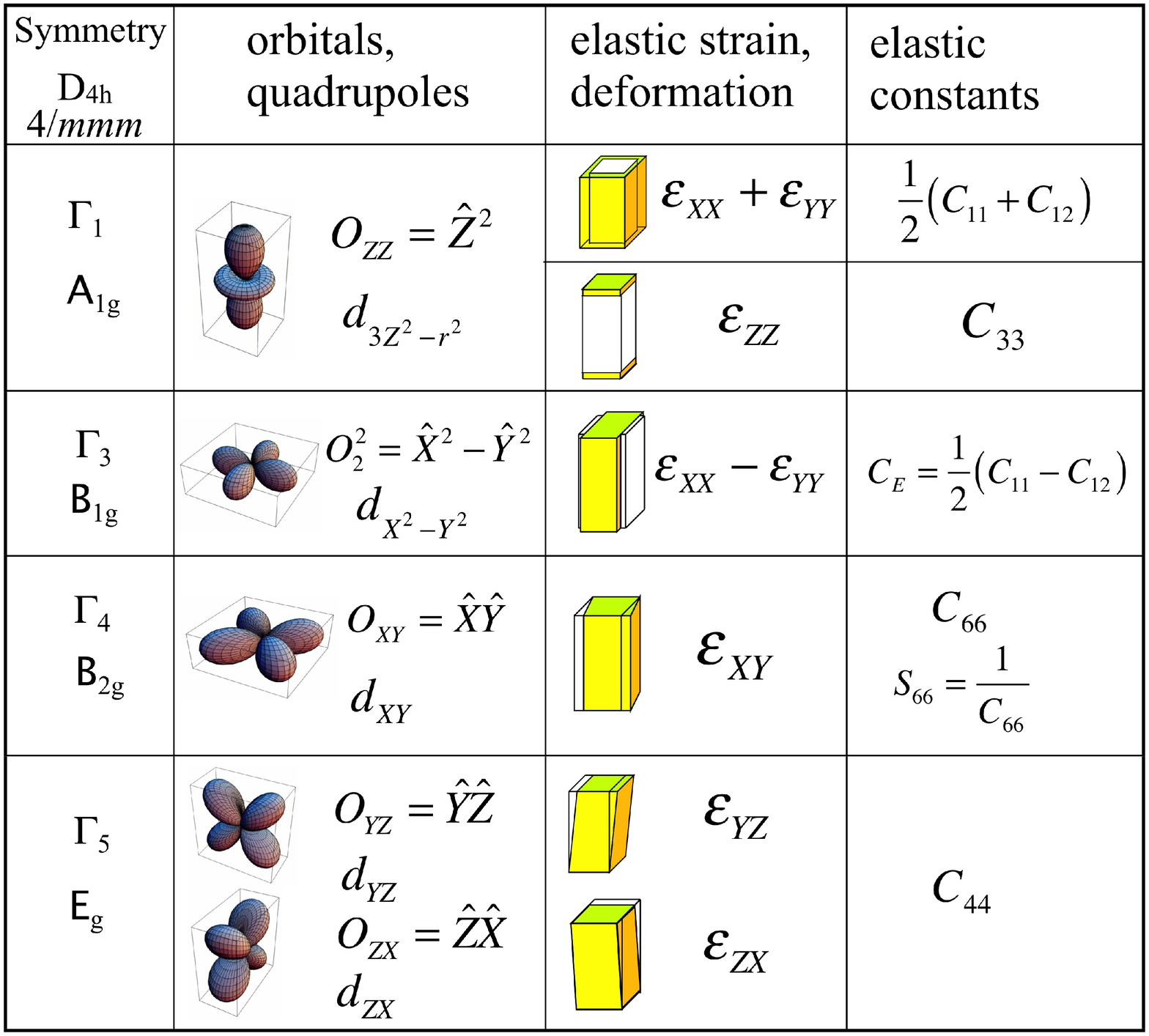}
\end{center}
\caption{(Color online)  Orbitals, quadrupoles, elastic strains, and their corresponding elastic constants classified into the irreducible representation of the point group $D_{4h}$.  
Figure cited from M. Yoshizawa {\it et al.}, Mod. Phys. Lett. B {\bf 26} (2012) 1230011. 
\copyright 2012, Modern Physics Letters B.}
\label{symmetry}
\end{figure}

It has been reported that some iron-based SCs show large elastic anomalies at low temperatures. 
Thin shape samples of polycrystalline LaFeAsO, BaFe$_{2}$As$_{2}$ and BaFe$_{1.84}$Co$_{0.16}$As$_{2}$ show elastic softening towards low temperatures.\cite{mcguire2008,fernandes2010}
Ultrasonic measurements for bulk samples have revealed that the elastic constant $C_{66}$ of Ba(Fe$_{1-x}$Co$_{x}$)$_{2}$As$_{2}$ (Ba122) shows a remarkably large elastic softening.\cite{goto2011,yoshizawa2012-1}
These anomalies are considered to be associated with the structural ordering of the neighboring phases.
In other words, tetragonal crystal symmetry is broken and changed to orthorhombic one in the neighboring phase.  

We have performed precise investigations of the elastic properties for Ba122, so far.
Then, we found that the elastic compliance $S_{66}$ (=1/$C_{66}$), a measure of structural fluctuation, behaves like the magnetic susceptibility near magnetic QCP, where it has been considered that spin fluctuation plays a relevant role for the emergence of superconductivity.
These experimental facts suggest the fluctuations with the same symmetry of the strains participate in the emergence of superconductivity.
These experimental studies have stimulated the interest of theoreticians. 
On the origin of the elastic anomaly, Fernandes {\it et al.} argued that the elastic anomalies are ascribed to nematic spin fluctuation.\cite{fernandes2010}
Kontani {\it et al.} proposed orbital origin of the large elastic anomalies.\cite{kontani2011,kontani2012}

These previous works suggest that the elastic constants are a suitable tool to investigate the fluctuations existing in iron-base SCs.
In our previous papers, it has been reported that the elastic constants of $C_{33}$ and $C_{66}$ of Ba122 system showed an elastic anomaly behavior near the QCP. 
According to our recent works, Fe(Se$_{1-x}$Te$_{x}$) shows elastic anomalies in elastic constants $C_{11}$, $C_{44}$, $(C_{11}-C_{12})/2$ in addition to $C_{66}$, and SrFe$_{2}$(As$_{1-x}$P$_{x}$)$_{2}$ exhibits larger elastic anomaly in $C_{44}$ than $C_{66}$.\cite{takezawa2014,horikoshi2017}
These works have implies the existence of diverse fluctuations in iron-based SCs.
It will be important object to investigate the relations between these fluctuations and their participation in the superconductivity.
In this article, we will report all the elastic constants of Ba(Fe$_{1-x}$Co$_{x}$)$_{2}$As$_{2}$ with eight Co-concentration to investigate the structural fluctuation with different symmetries in this system.
We also report the diverse fluctuations inherent in iron-based superconductors.

\section{Elastic constant measurement}
\subsection{Experimental Procedure}
\label{exp}
The elastic constant measurement was performed 
by an ultrasonic pulse-echo phase comparison method \cite{luethi2004} as a function of temperature where the temperature ranged from 5 to 300 K. The temperature was controlled using a cryostat mounted on a Gifford-McMahon (GM) cryocooler. 
To prevent the damage to the sample due to rapid changes in temperature, the rate of change in temperature was carefully controlled so as to be 10 K/h near $T_{\rm S}$ \cite{huang2008}.

\begin{figure*}
\begin{center}
\includegraphics[width=17cm]{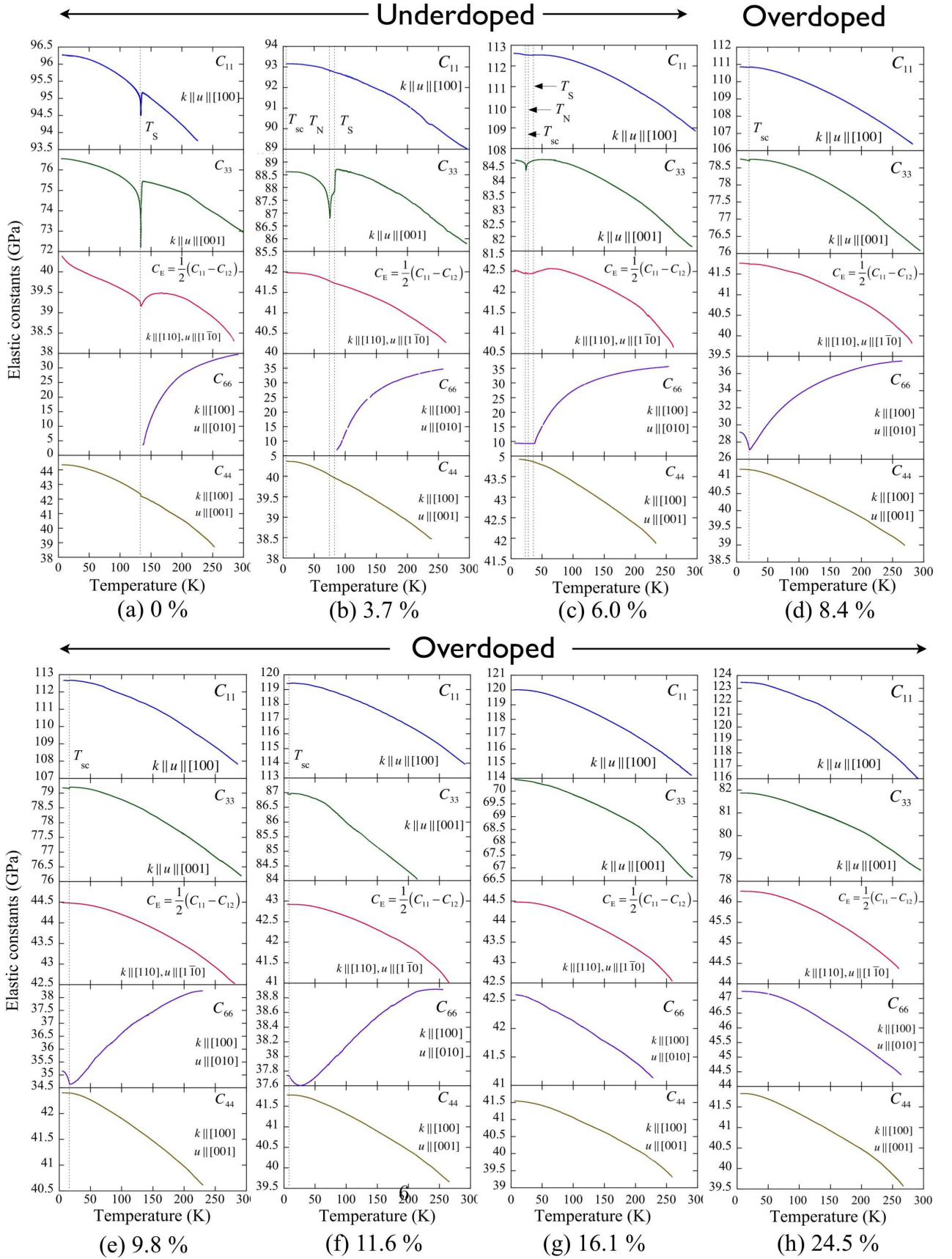}
\end{center}
\caption{(Color online) Temperature-dependence of the elastic stiffness constants $C_{11}$, $C_{33}$, $C_{\rm E}=(C_{11}-C_{12})/2$, $C_{66}$ and $C_{44}$ of Ba(Fe$_{1-x}$Co$_x$)$_2$As$_2$ with (a) $x$ = 0, (b) $x$ = 0.037, (c) $x$ = 0.060, (d) $x$ = 0.084, (e) $x$ = 0.098, (f) $x$ = 0.116, (g) $x$ = 0.161 and (h) $x$ = 0.245.}
\label{all}
\end{figure*}

Elastic stiffness was obtained by $C$ = $\rho v^2$, where $\rho$ is the density and $v$ is either the longitudinal or transverse sound velocity, $\rho$ was calculated by the lattice constant. 
Under the assumption of Vegard$^\prime$s law, the lattice constants of the $a\ (b)$- and $c$-axes are calculated by the data of $x$ = 0 and 0.1 of Ba(Fe$_{1-x}$Co$_x$)$_2$As$_2$ to be $a$ = $b$ = 0.39636 + 3.8981 $\times$ 10$^{-4}$$x$ (nm) and $c$ = 1.3022 - 0.0421$x$ (nm), respectively.\cite{sefat2008}
The corresponding longitudinal or transverse sound velocity is obtained by choosing the propagation and displacement directions.
In tetragonal crystal symmetry, we can measure six $C_{ij}$s; namely, $C_{11}$, $C_{33}$, $C_{12}$, $C_{13}$, $C_{44}$, and $C_{66}$.
The propagation and displacement directions of the sound velocity are respectively 
[100] and [100] for $C_{11}$, 
[001] and [001] for $C_{33}$, 
[100] and [010] for $C_{66}$, 
[100] and [001] for $C_{44}$, 
[110] and  $\left[ {1\bar 10} \right]$ for $\frac{1}{2}\left( {C_{11} - C_{12} } \right)$, 
and [110] and [110] for $C_{\rm{L}}  = \frac{1}{2}\left( {C_{11}  + C_{12}  + 2C_{66} } \right)$. 
Here, the $XYZ$ coordinate was defined by the unit cell of the $I4/mmm$ crystal structure \cite{rotter2008}, where the directions of $X$, $Y$ and $Z$ coincide with the principal axes of base-centered tetragonal lattice formed by Ba atoms, which is indicated in Fig. \ref{sample}. 
Absolute value of the sound velocity was obtained by the time interval of the echo train and the sample length, whose accuracy is within a few percent, dependent on the sample size. 

For the velocity measurements, ultrasound was emitted and detected using LiNbO$_{3}$ transducers. 
$Z$-cut LiNbO$_{3}$ with 100 ${\mu}$m thickness was used for longitudinal ultrasonic waves, and a 41$^\circ$ $X$-cut plate of LiNbO$_{3}$  with 100 ${\mu}$m thickness was used for transverse waves.
The fundamental frequencies of the longitudinal and transverse transducers were 33 and 19 MHz, respectively.
In this experiment, a third-higher harmonics of 114 and 64 MHz were applied to generate the longitudinal and transverse sound waves, respectively.
High-quality large single crystals of Ba(Fe$_{1-x}$Co$_x$)$_2$As$_2$ used in this work were grown by the self-flux method. 
Samples with eight Co-concentrations $x$ = 0, 0.037, 0.060, 0.084, 0.098, 0.116, 0.161, and 0.245 were prepared, and their corresponding compositions are shown in Fig. \ref{sample}.
The Co-concentration in the grown crystals was determined by energy-dispersive X-ray spectroscopy (EDS). 
Two samples were prepared for $x$ = 0.060, which are abbreviated as sample B and D.\cite{simayi2013}
The samples were cut into a rectangular shape, after determining their axis by X-ray Laue photograph. The samples have an average typical area of 3 $\times$ 3 mm$^2$ in the tetragonal ab $ab$\ ($XY$) cleavage plane, and thickness of 2 mm on the $c (Z)$-axis.

Recently, Kurihara {\it et al.} reported remarkable anomalies of ultrasonic attenuation coefficient for the same Co-doped Ba122 system. \cite{kurihara2017}
We have measured ultrasonic attenuation in addition to sound velocity, and observed ultrasonic attenuation anomalies in some elastic modes and samples.\cite{simayi2013}
The results on the ultrasonic attenuation are not included in this article, because we have not made the systematic study. 

\section{Experimental Results}
\label{result} 
\subsection{Underdoped region}
The measured temperature dependence of elastic constants $C_{ij}$ for all eight samples are summarized in Fig. \ref{all}. From the Fig. \ref{all} we know that the parent compound BaFe$_{2}$As$_{2}$ shows remarkable elastic softening associated with the structural phase transition $T_{\rm S}$ = 134 K in $C_{11}$, $C_{33}$ and $C_{66}$. $C_{44}$ increases with decreasing temperature and show a small anomaly at $T_{\rm S}$. $C_{\rm E}$ also shows remarkable anomaly at around $T_{\rm S}$.
In our previous paper, the large elastic softening in $C_{66}$ of Ba122 was presented. In that discussions, the temperature dependence of $C_{66}$ was characterized by a Jahn-Teller formula that corresponds to a Curie-Weiss expression for the magnetic susceptibility.\cite{yoshizawa2012-1}

\begin{figure}
\begin{center}
\includegraphics[width=8cm]{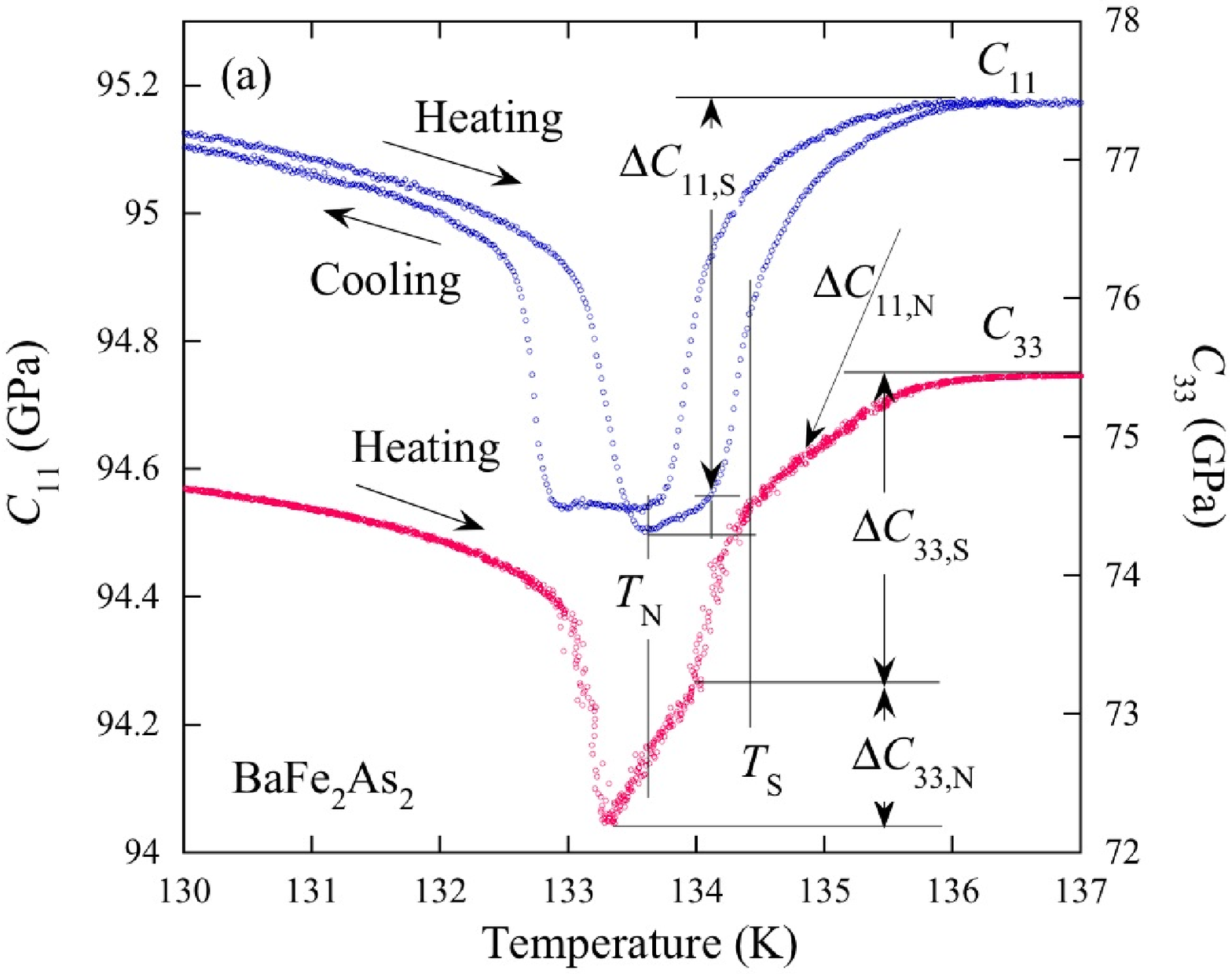}
\includegraphics[width=8cm]{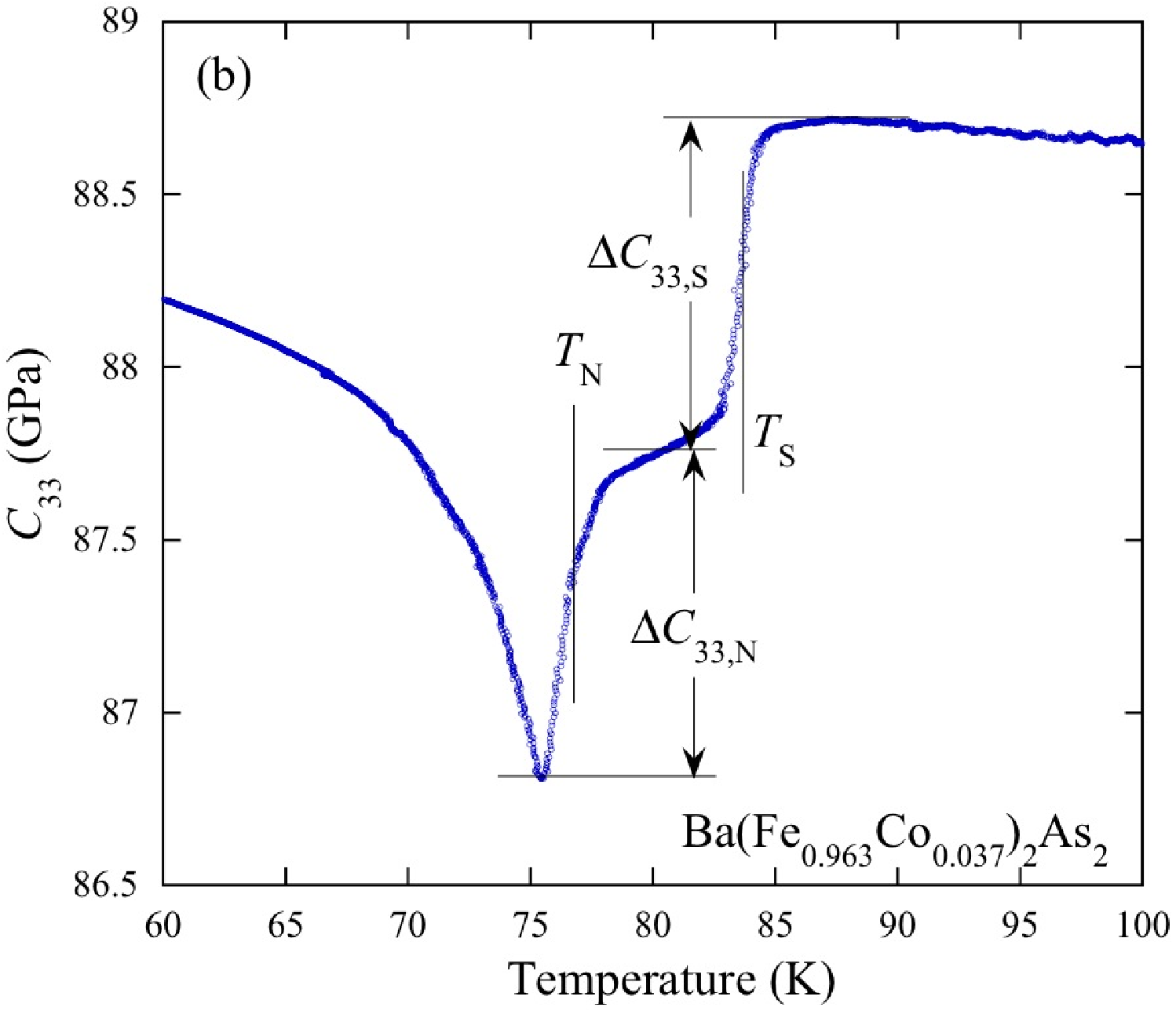}
\end{center}
\caption{(Color online) (a) Temperature dependence of $C_{11}$ and $C_{33}$ in BaFe$_{2}$As$_{2}$, and (b) $C_{33}$ in Ba(Fe$_{0.963}$Co$_{0.037}$)$_{2}$As$_{2}$.}
\label{C11}
\end{figure}

On the other hand, as illustrated in Fig. \ref{C11}, $C_{11}$(a) $C_{11}$ and $C_{33}$ show step-like temperature dependence, which comes from a magneto-elastic coupling between the structural and magnetic order parameters $\eta$ and the elastic strain $\varepsilon$ with the form of $\eta^{2}\varepsilon$. Also it can be seen from the Fig. \ref{C11}, $C_{11}$(a) that $C_{11}$ shows a hysteresis at low temperatures and it is dominated near the $T_{\rm N}$ and $T_{\rm S}$. This hysteresis appearing in $C_{11}$ implies that the structural transition is considered to be the first order. 
The elastic softening observed in $C_{11}$ for parent compound BaFe$_{2}$As$_{2}$ did not appeared in Co = 3.7\% sample.
However, $C_{33}$ shows elastic softening at two successive transition temperatures $T_{\rm N}$ and $T_{\rm S}$. The details are shown in Fig. \ref{C11}.

\subsection{Near optimal concentration}
\begin{figure}
\begin{center}
\includegraphics[width=8cm]{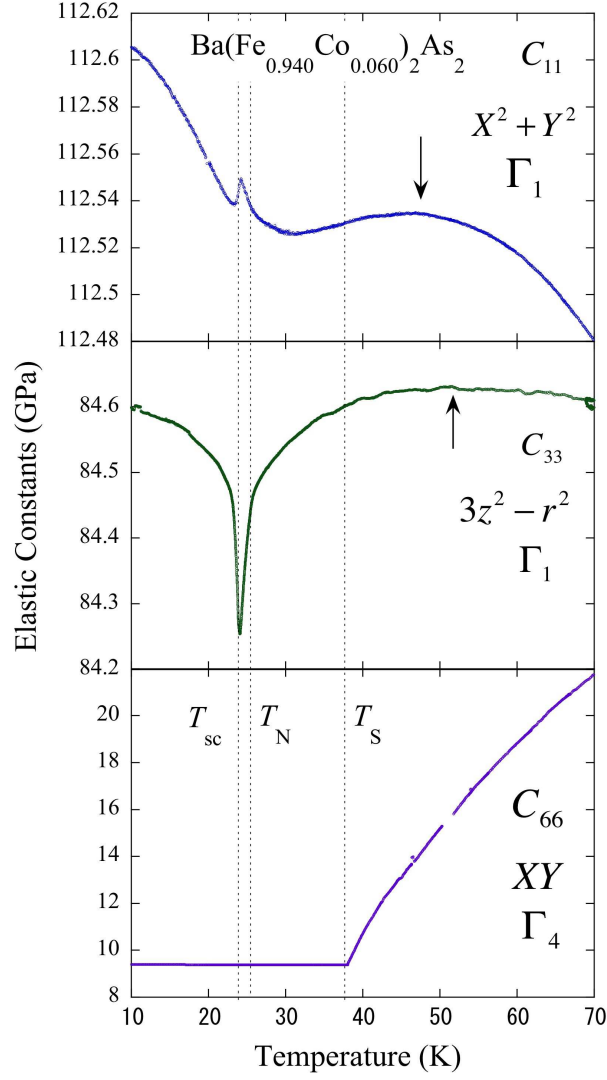}
\end{center}
\caption{(Color online)  Temperature dependence of $C_{11}$, $C_{33}$ and $C_{66}$. Their irreducible representations and bases are $\Gamma_{1}$ and $X^{2}+Y^{2}$ for $C_{11}$, $\Gamma_{1}$ and $3z^{2}-r^{2}$ for $C_{33}$, and $\Gamma_{4}$ and $XY$ for $C_{66}$. $T_{\rm S}$ is observed in sole $C_{66}$. On the other hand, $T_{\rm N}$ and $T_{\rm sc}$ are apparent in $C_{11}$ and $C_{33}$, which are followed by elastic softening from high temperatures. }
\label{C11C33C66}
\end{figure}

For Co = 6.0 \% sample which is near the optimal concentration, we have already reported precise temperature dependence of $C_{66}$ and $C_{33}$\cite{yoshizawa2012-1,simayi2013} and discussed in details about the elastic softening. From these results, it is remarkable that $C_{33}$ shows an elastic softening towards $T_{\rm N}$.
As shown in Fig. \ref{C11C33C66}, $C_{11}$ also shows a slight elastic softening at around 20 K in addition to the elastic softening in $C_{33}$. Superconducting transition is observed in both $C_{11}$ and $C_{33}$. 
It is interesting that $C_{11}$ shows a small anomaly at around $T_{\rm sc}$, however any anomaly observed at $T_{\rm N}$,  but $C_{33}$ shows a remarkable elastic softening towards $T_{\rm N}$ and $T_{\rm sc.}$.
Since the $C_{33}$ anomaly can be analyzed by Jahn-Teller formula, the elastic softening of $C_{33}$ would be ascribed to orbital fluctuations of inter-layer $O_{3z^{2}-r^{2}}$.
Thus, our results suggest the coexistence of different kinds of orbital fluctuations, in addition of $O_{xy}$ appearing in $C_{66}$.
Recently, a two-dome structure has been reported in the superconducting transition of 1111 system LaFeAsO$_{1-x}$H$_{x}$.\cite{fujiwara2013} 
On the origin of the superconductivity of this system, the important role of $O_{3z^{2}-r^{2}}$ orbital has been discussed.\cite{fujiwara2013,onari2014} 
By considering the above research results, we suppose that various fluctuations exist inherently in Ba122 system and play important role in the emergence of superconductivity. 

Here, we have to remark an important fact on the phase diagram. 
The difference between $T_{\rm S}$ and $T_{\rm N}$ is 1 K for BaFe$_{2}$As$_{2}$, 10 K for Co = 3.7 \% sample, and 12 K for Co = 6.0 \% sample.
The difference between $T_{\rm S}$ and $T_{\rm N}$ becomes larger with increasing the Co-concentration.
This behavior differs from the case of BaNi$_{2}$As$_{2}$, where $T_{\rm S}$ and $T_{\rm N}$ tends to merge by approaching the optimal concentration.
The experimental result of BaNi$_{2}$As$_{2}$ has been discussed from the point of avoided quantum criticality.\cite{lu2013}

\subsection{Overdoped region}
\begin{figure}
\begin{center}
\includegraphics[width=8cm]{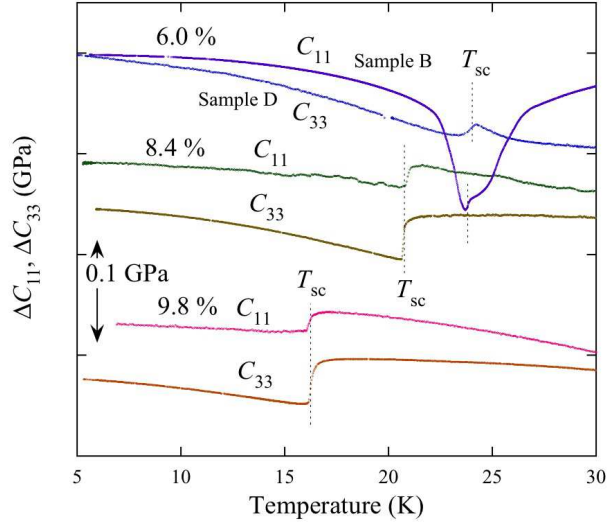}
\end{center}
\caption{(Color online)  Temperature dependence of $C_{11}$ and $C_{33}$ for 6.0 \%, 8.4 \% and 9.8 \% samples. $C_{11}$ data for 6.0 \% sample were taken from the sample B, and $C_{33}$ from the sample D, because the sample B does not show apparent anomaly at $T_{\rm sc}$ in $C_{33}$.}
\label{SCanomaly}
\end{figure}

Overdoped samples show rather monotonous temperature dependence in all elastic constants except $C_{66}$.
Elastic anomalies associated with $T_{\rm sc}$ were observed in all elastic constants.
In over-doped region, the elastic softening in $C_{66}$ above $T_{\rm sc}$ tends to disappear with the increase of Co-concentration.
As shown in Fig. \ref{SCanomaly}, $C_{11}$ and $C_{33}$ show step-like anomalies at $T_{\rm sc}$, which are caused by a magneto-strictive coupling between the superconducting order parameter and the elastic strain. 

\section{Discussion}
\subsection{Gr{\" u}neisen Parameter}
In this section, we will discuss about the Gr{\" u}neisen parameters for the structural transition temperature $T_{\rm S}$, the magnetic transition temperature $T_{\rm N}$, and the superconducting transition temperature $T_{\rm sc}$. 
Gr{\" u}neisen parameter is a measure of the interaction between order parameter and strain via magneto-strictive coupling, which can described by Eq. (\ref{definition}). 
We could obtained the absolute value of the Gr{\"u}neisen parameter $\Omega$ experimentally from the specific heat jump $\Delta C_{V}$ (we used $C_{p}$ instead of $C_{V}$) and the elastic constant jump $\Delta C$ at the transition temperature ($T_{\rm C}$) by using the following formula. 

\begin{equation}
{\Omega} =  - \frac{1}{{{T_{{\rm{C}}}}}}\frac{{\partial {T_{{\rm{C}}}}}}{{\partial {\varepsilon}}} 
\quad {\rm{and}}\quad \Delta {C} =  - \Omega^2\Delta {C_V}{T_{{\rm{C}}}}
\label{definition}
\end{equation}

\noindent Here, $\varepsilon$ is the elastic strain. 

\begin{figure}
\begin{center}
\includegraphics[width=8cm]{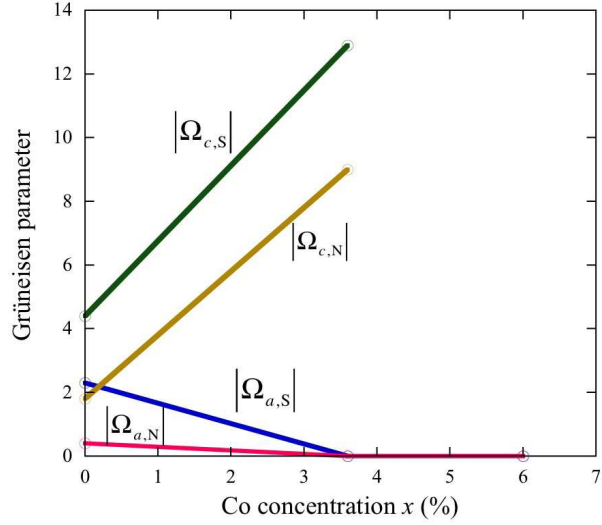}
\end{center}
\caption{(Color online)  Co-concentration dependence of Gr{\" u}neisen parameters $\Omega_{\rm S}$ and $\Omega_{\rm N}$.}
\label{Omega_TSTN}
\end{figure}

At first, we will calculate Gr{\" u}neisen parameter for $T_{\rm S}$ and $T_{\rm N}$ by using Eq. (\ref{def_TSTN}). 
The data used for the calculation are listed in Table I, and the Co-concentration dependence is shown in Fig.\ref{Omega_TSTN}.

\begin{subequations}
\begin{equation}
{\Omega _{a,\,{\rm S}}} =  - \frac{1}{{{T_{\rm{S}}}}}\frac{{\partial {T_{\rm{S}}}}}{{\partial {\varepsilon _{XX}}}}\quad {\rm{and}}\quad \Delta {C_{11,\,{\rm{S}}}} =  - \Omega _{a,\,{\rm S}}^2\Delta {C_{p,\,{\rm{S}}}}{T_{\rm{S}}},
\end{equation}
\begin{equation}
{\Omega _{c,\,{\rm S}}} =  - \frac{1}{{{T_{\rm{S}}}}}\frac{{\partial {T_{\rm{S}}}}}{{\partial {\varepsilon _{ZZ}}}}\quad {\rm{and}}\quad \Delta {C_{33,\,{\rm{S}}}} =  - \Omega _{c,\,{\rm S}}^2\Delta {C_{p,\,{\rm{S}}}}{T_{\rm{S}}},
\end{equation}
\begin{equation}
{\Omega _{a,\,{\rm N}}} =  - \frac{1}{{{T_{\rm{S}}}}}\frac{{\partial {T_{\rm{N}}}}}{{\partial {\varepsilon _{XX}}}}\quad {\rm{and}}\quad \Delta {C_{11,\,{\rm{N}}}} =  - \Omega _{a,\,{\rm N}}^2\Delta {C_{p,\,{\rm{N}}}}{T_{\rm{N}}},
\end{equation}
\begin{equation}
{\Omega _{c,\,{\rm N}}} =  - \frac{1}{{{T_{\rm{N}}}}}\frac{{\partial {T_{\rm{N}}}}}{{\partial {\varepsilon _{ZZ}}}}\quad {\rm{and}}\quad \Delta {C_{33,\,{\rm{N}}}} =  - \Omega _{c,\,{\rm N}}^2\Delta {C_{p,\,{\rm{N}}}}{T_{\rm{N}}}.
\end{equation}
\label{def_TSTN}
\end{subequations}

\begin{table}
\caption{Anomalous part of $C_{11}$ and $C_{33}$ at $T_{\rm S}$ ($\Delta C_{11,\, {\rm S}}$ and $\Delta C_{33,\, {\rm S}}$), specific heat anomaly at $T_{\rm S}$ ($\Delta C_{p,\, {\rm S}}$), and the absolute values of calculated Gr{\" u}neisen parameters $\left| {{\Omega _{a,\,{\rm{S}}}}} \right|$ and $\left| {{\Omega _{c,\,{\rm{S}}}}} \right|$ for BaFe$_{2}$As$_{2}$, the 3.7\%, and 6.0\%-doped samples. 
Anomalous part of $C_{11}$ and $C_{33}$ at $T_{\rm N}$ ($\Delta C_{11,\, {\rm N}}$ and $\Delta C_{33,\, {\rm N}}$), specific heat anomaly at $T_{\rm N}$ ($\Delta C_{p,\, {\rm N}}$), and the absolute values of calculated Gr{\" u}neisen parameters $\left| {{\Omega _{a,\,{\rm{N}}}}} \right|$ and $\left| {{\Omega _{c,\,{\rm{N}}}}} \right|$ for BaFe$_{2}$As$_{2}$, the 3.7\%-, and 6.0\%-doped samples. 
Specific heat data were taken from Simayi {\it et al}.\cite{simayi2013} and unpublished data.
\noindent*1) No precise measurement, but it would be concluded to be very small from rough measurements.
}
\centering
\begin{tabular}{p{0.35\linewidth}p{0.1\linewidth}p{0.1\linewidth}p{0.1\linewidth}}
\hline
$x$-Co (\%)    & 0 &  3.7 & 6.0   \\
\hline
\hline
$T_{\rm S}$ (K)    & 134.4 & 83.7 & 37.5  \\
\hline
$\Delta C_{11,\, {\rm S}}$ (GPa)   & -0.62 & 0 & 0  \\
\hline
$\Delta C_{33,\, {\rm S}}$ (GPa)   & -2.27 & -0.95 & 0   \\
\hline
$\Delta C_{p,\, {\rm S}}$ (mJ/mol$\cdot$K) & 3.0 & 0.6 & *1 \\
\hline
$\left| {{\Omega _{a,\,{\rm{S}}}}} \right|$ & 2.3 & 0 & --- \\
\hline
$\left| {{\Omega _{c,\,{\rm{S}}}}} \right|$ & 4.4 & 12.9 & --- \\
\hline
\hline
$T_{\rm N}$ (K)    & 133.6 & 76.8 & 25.7  \\
\hline
$\Delta C_{11,\, {\rm N}}$ (GPa)   & -0.057 & 0 & 0  \\
\hline
$\Delta C_{33,\, {\rm N}}$ (GPa)   & -1.00 & -0.93 & 0.21   \\
\hline
$\Delta C_{p,\, {\rm N}}$ (mJ/mol$\cdot$K) & 8.8 & 1.55 & $\approx$0 \\
\hline
$\left| {{\Omega _{a,\,{\rm{N}}}}} \right|$ & 0.4 & 0 & --- \\
\hline
$\left| {{\Omega _{c,\,{\rm{N}}}}} \right|$ & 1.8 & 9.0 & --- \\
\hline
\end{tabular}
\end{table}

$\Omega_{c,{\rm S}}$ and $\Omega_{c,{\rm N}}$ increase towards the QCP,
while $\Omega_{a,{\rm S}}$ and $\Omega_{a,{\rm S}}$ tend to be zero as approaching QCP.
For the Co = 6.0 \% sample, no specific heat anomaly was found at $T_{\rm N}$ and $T_{\rm S}$, because the temperature was roughly scanned near these transition temperatures.
We suppose, however, they are very small.
It may imply that both Gr{\" u}neisen parameter $\Omega_{c,\, {\rm S}}$ and $\Omega_{c,\, {\rm N}}$ are considered to be quite large for 6.0 \%, and would show divergent behavior towards QCP.
According to the recent works, the Gr{\" u}neisen parameter is expected to be divergent near the QCP.\cite{zhu2003,kuechler2006}
Although the iron-based superconductors show in-plane structural and magnetic orders, however the calculated Gr{\" u}neisen parameters in this study related to the inter-plan direction show a relevant Gr{\" u}neisen parameters. 
Next, Gr{\" u}neisen parameters for $T_{\rm sc}$ were evaluated by the same manner by using Eq.(\ref{def_Tsc}) and the data listed in Table II.

\begin{subequations}
\begin{equation}
{\Omega _a} =  - \frac{1}{{{T_{{\rm{sc}}}}}}\frac{{\partial {T_{{\rm{sc}}}}}}{{\partial {\varepsilon _{XX}}}}\quad {\rm{and}}\quad \Delta {C_{11,\,{\rm{sc}}}} =  - \Omega _a^2\Delta {C_{p,\,{\rm{sc}}}}{T_{{\rm{sc}}}},
\end{equation}
\begin{equation}
{\Omega _c} =  - \frac{1}{{{T_{{\rm{sc}}}}}}\frac{{\partial {T_{{\rm{sc}}}}}}{{\partial {\varepsilon _{ZZ}}}}\quad {\rm{and}}\quad \Delta {C_{33,\,{\rm{sc}}}} =  - \Omega _c^2\Delta {C_{p,\,{\rm{sc}}}}{T_{{\rm{sc}}}}.
\end{equation}
\label{def_Tsc}
\end{subequations}

\noindent The values of $\Omega_{c}$ were already reported in the previous paper.\cite{simayi2013}
In addition to $\Omega_{c}$, we evaluated $\Omega_{a}$ in this work.
We cannot determine the sign of $\Omega_{a}$ and $\Omega_{c}$ from the elastic constant measurement. 
Some hypothesis or careful consideration will be needed to determine their sign.
We will discuss later on this point.

Here, we will compare our results with those of previous thermal expansion measurements.
For comparison, we should convert our results as a function of the uniaxial strain dependence $dT_{\rm sc}/d\varepsilon_{\rm i}$ to that of the uniaxial pressure dependence of $T_{\rm sc}$ by using the following formulas:

\begin{subequations}
\begin{equation}
\frac{{d{T_{{\rm{sc}}}}}}{{d{\varepsilon _{XX}}}} =  - \left( {{C_{11}} + {C_{12}}} \right)\frac{{d{T_{{\rm{sc}}}}}}{{d{p_a}}} - {C_{13}}\frac{{d{T_{{\rm{sc}}}}}}{{d{p_c}}},
\end{equation}
\begin{equation}
\frac{dT_{\rm sc}}{d\varepsilon_{ZZ}}=-C_{\rm 33}\frac{dT_{\rm sc}}{dp_{c}}-2C_{13}\frac{dT_{\rm sc}}{dp_{a}}.
\end{equation}
\label{convert}
\end{subequations}

\noindent The values of 109, 79, and 29 GPa were used for $C_{11}$, $C_{33}$, and $C_{12}$ respectively. 
In this calculation, it was assumed that $C_{13}$ is the same value as $C_{12}$ because $C_{13}$ was not measured and is unknown.
Bud$^{'}$ko {\it et al.} reported $dT_{\rm sc}/dp_{i}$ for 3.8\% and 7.4\%, and Hardy {\it et al.} reported $dT_{\rm sc}/dp_{i}$ for 8.0\%.
We compare these values of $\frac{{d{T_{{\rm{sc}}}}}}{{d{\varepsilon _{XX}}}}$ obtained by our measurements with those calculated by using the reported values of $\frac{{d{T_{{\rm{sc}}}}}}{{d{p_a}}}$ and $\frac{{d{T_{{\rm{sc}}}}}}{{d{p_c}}}$.  
Hardy {\it et al.} obtained $dT_{\rm sc}/dp_{\rm a}=3.1(1)$ K/GPa and $dT_{\rm sc}/dp_{c}=-7.0(2)$ K/GPa for 8\% doped sample from the thermal expansion measurement.\cite{hardy2009}
The value of $dT_{\rm sc}/dp_{c}$ is comparable to that by Nakashima {\it et al.} of -13 K/GPa for 8\% doped sample.\cite{nakashima2010}
These values give the $\Omega_{a}$ and $dT_{\rm sc}/d\varepsilon_{XX}$ to be 10.0 and -219 K, 
and $\Omega_{c}$ and $dT_{\rm sc}/d\varepsilon_{ZZ}$ to be -16.7 and 365 K, respectively, for 8\% doped sample.
These values are consistent with our results of $\left| {{\Omega _a}} \right|$ = 11.2 and $\left| {{\Omega _c}} \right|$ = 16.2 for the 8.4 \% doped sample.

On the other hand, Bud$^{'}$ko {\it et al.} reported $dT_{\rm sc}/dp_{a}=-4.1$ K/kbar and $dT_{\rm sc}/dp_{c}=1.7$ K/kbar for 3.8\% doped sample, $dT_{\rm sc}/dp_{a}=0.3$ K/kbar and $dT_{\rm sc}/dp_{\rm c}=-2.6$ K/kbar for 7.4\% doped sample \cite{budko2009-2}. 
Gr{\" u}neisen parameters for the 7.4 \% sample are $\Omega_{a}$ = -14.2 and $\Omega_{c}$ = -91.6, which are remarkably different from our results.
Origin of the inconsistency between our results and Bud'ko's ones is an enigma.

Next, we have to pay our attention to the sign of the Gr{\" u}neisen parameter.
We try to determine their sign from the information of thermal expansion data.
We were aware of some tendency in the $dT_{\rm sc}/dp_{a}$ and $dT_{\rm sc}/dp_{c}$ as a function of Co-concentration in the previous thermal expansion measurements.
$dT_{\rm sc}/dp_{c}$ is positive for underdoped samples, and negative for the overdoped doped samples.\cite{budko2009-2}
$dT_{\rm sc}/dp_{a}$ has an opposite sign of $dT_{\rm sc}/dp_{c}$.\cite{hardy2009} 
The decrease of $T_{\rm sc}$ was reported by the uniaxial pressure along $c$-axis for an overdoped sample.\cite{nakashima2010} 
These results suggest that the sign of $\Omega_{a}$ and $\Omega_{c}$ are negative and positive, respectively, in the underdoped region, and vise versa in overdoped region. 
Therefore, we assumed that sign of the Gr{\" u}neisen parameters follows along a general tendency observed in the previous works, and listed in Table II.

\begin{figure}
\begin{center}
\includegraphics[width=8cm]{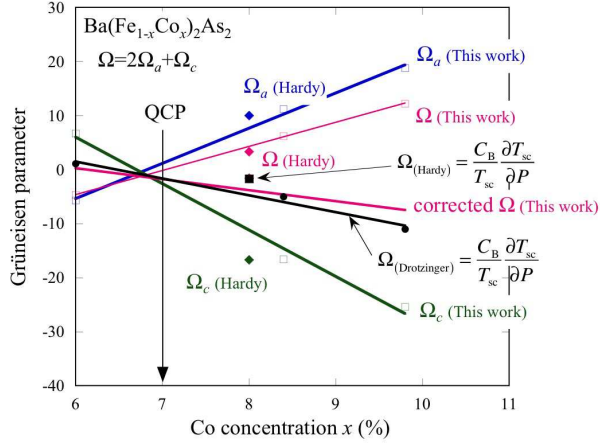}
\end{center}
\caption{(Color online)  Co-concentration dependence of Gr{\" u}neisen parameter $\Omega_{\rm sc}$.}
\label{Omega}
\end{figure}

Figure \ref{Omega} shows the Co-concentration dependence of $\Omega_{\rm sc}$. The calculated Gr{\" u}neisen parameters in this study have been plotted with the data from Hardy {\it et al.} and Drotziger {\it et al}.\cite{hardy2009,drotziger2010}
To compare our results with those of Drotziger {\it et al.}, we evaluated ${{\partial {T_{{\rm{sc}}}}} \mathord{\left/
 {\vphantom {{\partial {T_{{\rm{sc}}}}} {\partial P}}} \right.
 \kern-\nulldelimiterspace} {\partial P}}$ 
 from their article, and obtained by the formula of 
 $\left( {{{{C_{\rm{B}}}} \mathord{\left/
 {\vphantom {{{C_{\rm{B}}}} {{T_{{\rm{sc}}}}}}} \right.
 \kern-\nulldelimiterspace} {{T_{{\rm{sc}}}}}}} \right)\left( {{{\partial {T_{{\rm{sc}}}}} \mathord{\left/
 {\vphantom {{\partial {T_{{\rm{sc}}}}} {\partial P}}} \right.
 \kern-\nulldelimiterspace} {\partial P}}} \right)$.
 Here, $C_{\rm B}$ is the bulk modulus, which was evaluated to be 44 GPa by using 
 
 \begin{equation}
 {C_{\rm{B}}} = \frac{{\left( {{C_{11}} + {C_{12}}} \right){C_{33}} - 2C_{13}^2}}{{{C_{11}} + 2{C_{33}} + {C_{12}} - 4{C_{13}}}},
 \label{Bulk_modulus}
 \end{equation}

\noindent under the assumption of $C_{13} = C_{12}$.
In the case of Hardy's result, the sign of $\Omega$ obtained by the formula $\Omega = 2\Omega_{a}+\Omega_{c}$ is positive, and that obtained by 
$\left( {{{{C_{\rm{B}}}} \mathord{\left/
 {\vphantom {{{C_{\rm{B}}}} {{T_{{\rm{sc}}}}}}} \right.
 \kern-\nulldelimiterspace} {{T_{{\rm{sc}}}}}}} \right)\left( {{{\partial {T_{{\rm{sc}}}}} \mathord{\left/
 {\vphantom {{\partial {T_{{\rm{sc}}}}} {\partial P}}} \right.
 \kern-\nulldelimiterspace} {\partial P}}} \right)$
 is negative.
 This inconsistency was considered to be caused by the values of the elastic constants used in Eqs. (3) and (4).
 We consider that the sign of the Gr\"{u}neisen parameters alternate with the opposite one near the QCP.
If the hydrostatic pressure was applied, the sign of 
${{\partial {T_{{\rm{sc}}}}} \mathord{\left/
 {\vphantom {{\partial {T_{{\rm{sc}}}}} {\partial P}}} \right.
 \kern-\nulldelimiterspace} {\partial P}}$
is positive for the underdoped samples and negative for the overdoped samples. 
It would be reasonable, because $T_{\rm sc}$ shows a maximum near the QCP, which results that $T_{\rm sc}$ is not influenced by the change of any external parameter.
This means that the bulk Gr\"{u}neisen parameter $\Omega$ should be approximately zero near the QCP.
In our original data, $\Omega$ is negative for the underdoped region and positive for the over-doped region, and it increases with the increasing of Co-concentration, it would be somewhat unreasonable.
As the same reason as the inconsistency in the Hardy's result, if we try to change the values of $\Omega_{a}$ and $\Omega_{c}$ by multiplying some coefficients; namely, $0.7\Omega_{a}$, $1.3\Omega_{c}$ and $\Omega_{\rm corrected} = 0.7\Omega_{a}+ 1.3\Omega_{c}$, $\Omega_{\rm corrected}$ takes positive sign in underdoped region and negative sign in overdosed region and shows similar behavior to Drotziger {\it et al}.\cite{drotziger2010}
We plotted these Gr{\" u}neisen parameters as a function of Co-concentration in Fig. \ref{Omega}.

\begin{table}
\caption{$T_{\rm sc}$, $\Delta C_{33}$, $\Delta C_{p}/T_{\rm sc}$, $dT_{\rm sc}/dp_{c}$, calculated $dT_{\rm sc}/{d\varepsilon_{ZZ}}$, and  $\Omega$ values for the 6.0\%-, 8.4\%-, and 9.8\%-doped samples. Specific heat data were taken from Simayi {\it et al}.\cite{simayi2013} and unpublished data.}
\centering
\begin{tabular}{p{0.35\linewidth}p{0.1\linewidth}p{0.1\linewidth}p{0.1\linewidth}}
\hline
$x$-Co (\%)    & 6.0 &  8.4 & 9.8   \\
\hline
$T_{\rm sc}$ (K)    & 24.0 & 20.6 & 16.7   \\
\hline
$\Delta C_{11,\, {\rm sc}}$ (10$^{-2}$GPa)   & -1.0 & -2.0 & -2.4  \\
\hline
$\Delta C_{33,\, {\rm sc}}$ (10$^{-2}$GPa)   & -1.4 & -4.2 & -4.4   \\
\hline
$\Delta C_{p,\, {\rm sc}}$ (mJ/mol$\cdot$K) & 792 & 474 & 251 \\
\hline
$\Omega_{a}$ & -5.7 & 11.2 & 18.8 \\
\hline
$\Omega_{c}$ & 6.7 & -16.2 & -25.2 \\
\hline
$\Omega$ & -4.7 & 6.2 & 12.4 \\
\hline
\end{tabular}
\end{table}

In this section, the Gr{\" u}neisen parameters have been calculated for for $T_{\rm S}$, $T_{\rm N}$ and $T_{\rm sc}$ as a function of Co-concentration.
$T_{\rm S}$ and $T_{\rm N}$ are not affected by in-plane elongation and compression, but did not affected by in-plane elongation and compression.
These findings in this work is very striking, because it has been believed that structural and magnetic orders are quasi-2-dimensional and sensitive to in-plane deformation.
On the other hand, from the above results we know that in-plane contraction and inter-plane elongation enhance $T_{\rm SC}$.
In particular, the $c$-axis elongation stabilizes the superconductivity. 
The same deformation also promote the structural and magnetic orders. 
These facts suggest that the existence of certain correlation between the structural and magnetic orders and superconductivity. 
$\Gamma_{1}$ fluctuation appearing in $C_{33}$ plays an important role in the emergence of superconductivity in addition to in-plane $C_{66}$ fluctuations.

\subsection{Elastic anomaly associated with superconductivity}

\begin{figure}
\begin{center}
\includegraphics[width=8cm]{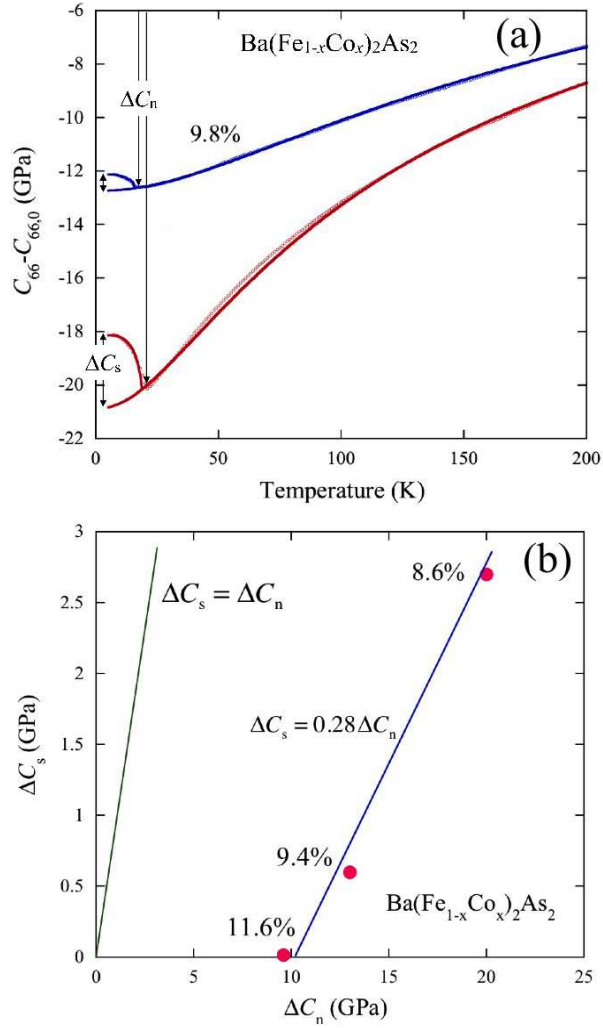}
\end{center}
\caption{(Color online)  (a) Temperature dependence of $C_{66}$ and. Bold lines above $T_{\rm sc}$ are the calculation taken from the previous study.\protect\cite{yoshizawa2012-1}. The curves below $T_{\rm sc}$ are the guides for eyes. (b) Anomaly associated with superconductivity $\Delta C_{\rm s}$ as a function of the elastic anomaly in the normal phase $\Delta C_{\rm n}$ (= $C_{66} - C_{66,0}$ at $T_{\rm sc}$).}
\label{total_symmetry}
\end{figure}

In this subsection, we will focus our attention on 
the elastic anomaly associated with the superconductivity.
It has been reported that the elastic anomaly in $C_{66}$ at $T_{\rm sc}$ is quite different between underdoped and overdoped regions.
In the under-doped region, $C_{66}$ shows an elastic hardening at $T_{\rm N}$, and shows softening at $T_{\rm sc}$.\cite{yoshizawa2012-1} 
On the other hand, $C_{66}$ increases with an up-turn at $T_{\rm sc}$ followed by a large elastic softening.\cite{goto2011,yoshizawa2012-1}
This behavior ascribed to the phase transition in tetragonal phase (overdoped case) and orthorhombic phase (underdoped case).
In other words, we suppose that it would be a signature that superconductivity of both cases are qualitatively different from each other.

For the overdoped region,
Fig. \ref{total_symmetry}(a) is the temperature dependence of $C_{66}$ for 8.4 \% and 9.8 \% samples.
Bold lines above $T_{\rm sc}$ are the theoretical fit by using the following equations, taking the band contribution into account, which were precisely reported elsewhere.\cite{yoshizawa2012-1}

The elastic anomaly associated with band electrons is described by the following charge susceptibility. 

\begin{equation}
\label{bandChi}
\chi _{\rm_{S}} =  - \int {dE\,N\left( E \right)} \frac{{\partial f\left( E \right)}}{{\partial E}}
\end{equation}

\noindent where $f$ and $N$ the Fermi-Dirac function 
and the density of states.

The curves below $T_{\rm sc}$ in Fig. \ref{total_symmetry}(a) are guides for eyes.
According to Eq.(\ref{bandChi}), the elastic anomaly by the band electron is considered to vanish associated with a gap opening in the superconducting phase, because $N_{0}$ is zero below $T_{\rm sc}$.
This makes $C_{66}$ increase below $T_{\rm sc}$.
From this consideration, the amount of elastic anomaly above $T_{\rm sc}$ ($\Delta C_{\rm{n}}$) is expected to be equal to that below $T_{\rm sc}$ ($-\Delta C_{\rm{s}}$) at $T$ = 0 K, because the density of states at Fermi energy is zero for a full-gap SCs.

In Fig. \ref{total_symmetry}(b) shows the plotted $\Delta C_{\rm{n}}$ as a function of $-\Delta C_{\rm{s}}$ for 8.4 \%, 9.8 \% and 11.6 \% samples.
We found two peculiar behaviors from the figure.
First, the relation of $\Delta C_{\rm s} = -\Delta C_{\rm n}$ does not hold.
The ratio of $\Delta C_{\rm s}$ against $-\Delta C_{\rm n}$ was evaluated to be 0.28 from the tendency of the three samples, which is indicated by the straight line in Fig. \ref{total_symmetry}(b).
If the elastic anomaly in the normal phase would be contributed solely by the band, than the band with a gap opened is 28 \% of the band responsible for the elastic softening.
The rest of 72 \% is considered to be gapless, or does not participate in the superconductivity.
Second, $\Delta C_{\rm s}$ vanishes more steeply than $-\Delta C_{\rm n}$ with increasing Co-concentration.
$\Delta C_{\rm s}$ reaches to zero when $-\Delta C_{\rm n}$ = 10 GPa.
The second finding is that $\Delta C_{\rm s}$ decreases rapidly and almost zero for 11.6 \% sample, although its $T_{\rm sc}$ = 10.5 K. 

These phenomena are very peculiar, and the origin is an enigma.
However, it should be remark that these discussions in this work are considered to be valid for typical $s_{++}$ full-gap SCs.
Therefore, the origin of these behavior should be carefully discussed, and will remain as a future task.

\begin{figure}
\begin{center}
\includegraphics[width=8cm]{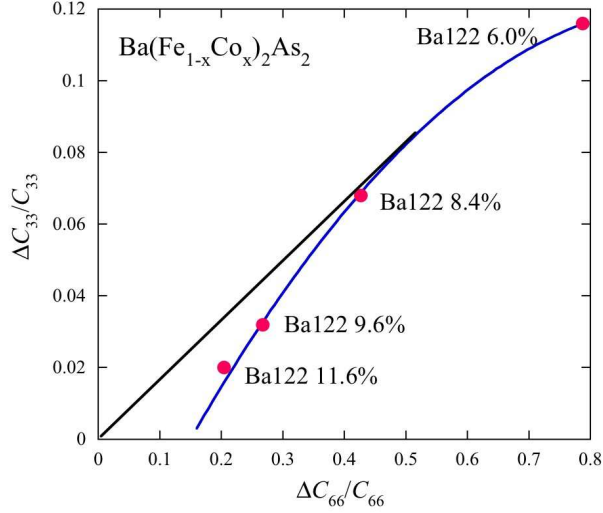}
\end{center}
\caption{(Color online)  Temperature dependence of $C_{11}$, $C_{33}$ and $C_{66}$.}
\label{C66vsC33}
\end{figure}

\section{Conclusion}
In this work, total 40 elastic constants with five elastic modes of $C_{11}$, $C_{33}$, $C_{\rm E} = (C_{11}-C_{12})/2$, $C_{44}$ and $C_{66}$ for Ba(Fe$_{1-x}$Co$_{x}$)$_{2}$As$_{2}$ (where x = x = 0, 3.7, 6.0, 8.4, 9.8, 11.6, 16,1 and 24.5) single crystalline samples have been presented.
We focused our attention on the elastic anomalies in $C_{11}$ and $C_{33}$ modes, and discussed Gr{\" u}neisen parameters of this system.
Although, some assumption for the determination of the sign for Gr{\" u}neisen parameters were adopted, it would be considered that Gr{\" u}neisen parameters of Ba122 system are very anisotropic.
t would be accepted that the bulk Gr{\" u}neisen $\Omega$ is zero near optimal concentration.
Although our results surely indicate that both $\Omega_{a}$ and $\Omega_{c}$ are very small near the optimal concentration, they develop differently with leaving from the optimal concentration.
The bulk $\Omega$ keeps nearly zero as a consequence of cancellation of $\Omega_{a}$ and $\Omega_{c}$ with an opposite sign.
Owing to this property, $T_{\rm sc}$ could be stable against hydrostatic pressure, however the $T_{\rm sc}$ is expected to be enhanced by uniaxial pressure. 
Higher $T_{\rm sc}$ would be realized by contraction of $ab$-plane and expansion along $c$-direction by chemical treatment of replacing atoms.

This suggestion can be replaced by other word from the viewpoint of relevant role of structural fluctuation along $c$-axis, because $c$-axis elongation is preferable for $3z^{2}-r^{2}$ orbital.
In our previous paper, we pointed out anomalous softening in $C_{33}$ near the optimal concentration.\cite{simayi2013}
Figure \ref{C11C33C66} shows that $C_{11}$ anomaly exists in addition to $C_{33}$.
$C_{11}$ and $C_{33}$ softening represent C$_{4}$ fluctuation.
Figure \ref{C66vsC33} shows the amount of $C_{33}$ softening as a function of the amount of $C_{66}$ softening.
There exists a correlation between the amount of $C_{33}$ and $C_{66}$ anomalies.
While $C_{66}$ expresses C$_{2}$ fluctuation, $C_{33}$ is C$_{4}$ fluctuation.
Two types of fluctuations seem to cooperate near the QCP.
It would be concluded that the superconductivity with high $T_{\rm sc}$ of this system is brought about by the collaboration of C$_{2}$ fluctuation and C$_{4}$.

\section*{Acknowledgments}
We would like to thank T. Kowata for his help in determining the atomic composition of the samples by using EDS.
This work was supported by the Transformative Research-project on Iron Pnictides of the Japan Science and Technology Agency,  JSPS KAKENHI Grant Number 24500350 and JP15H05883 (J-Physics).
\subsection{A subsection}

More text.

\end{document}